\documentstyle[prl,aps]{revtex}
\input{epsf}

\begin{document}
\title{Electronic structure without exchange?}
\author{T. Jarlborg}
\address{DPMC, University of Geneva, 
24 Quai Ernest-Ansermet, CH-1211 Gen\`eve 4, Switzerland}
\baselineskip 5mm
\maketitle


\begin{abstract}
The correlation holes for densities of equal and opposite spin around a test electron
are determined from the Schr\"{o}dinger equation with proper boundary conditions. 
 The traditional
"exchange" term follows from the boundary condition which respects a spacial 
exclusion principle for equal spins.
The resulting potential compares reasonably well with standard local
density potentials and should simplify extensions towards non-local effects. 

\end{abstract}


 The Pauli principle excludes multiple occupancy of spin orbitals. When the wave function is
a Slater determinant, it gives rise to the concept of exchange, i.e. when two electron
indicies in the Coulomb integral are exchanged there will be a perfect annulation of the direct
Coulomb term for multiple occupancy \cite{sla}. An exchange hole is created around a certain
electron, excluding other electrons of the same spin to come too close. When applied to
free electron wave functions, this exchange hole has a well-known form, and 
the resulting exchange potential is in the 
Hartree-Fock-Slater approximation (HFS) applied
to atoms, molecules and solids, which in general are not free electron like. In this way it
is possible to reduce the difficult N-body problem to a 1-particle like Schr\"{o}dinger
equation:
\begin{equation}
\label{eq:sch}
-\nabla^2 \Psi_i(r) + (V_{ext}(r) + \int \frac{\rho(r')}{\mid r - r' \mid}d^3r' + 
\mu_{xc}(r))\Psi_i(r) = \epsilon_i\Psi_i(r)
\end{equation}
where $\Psi_i(r)$ is the wave function for electron state $i$ with energy $\epsilon_i$,
and $\rho(r)$ is the electron 
density. The external potential $V_{ext}(r)$ is the attractive potential due to all
nuclear charges and the exchange-correlation potential is reduced in a local form, $\mu_{xc}(r)$.
 The 
density functional (DF) theory in the local density approximation (LDA) showed that
this is correct in variational calculations with the electron density as a variable \cite{lda}.
The exchange hole still serves as a valid pair correlation function between two electrons
of the same spin, but the resulting exchange potential is in the LDA Kohn-Sham (KS) potential
2/3 of that in HFS. Correlation is an additional effect that stems from the fact that all electrons
repel each other, so that due to the correlated movement of the electrons there will
be a reduced Coulomb term in the in the DF-type Schr\"{o}dinger equation for a one-particle
potential \cite{sstl}. Similarly as the exchange hole for equal spin, there will be a correlation hole
also for opposite spin electron density around an electron, but it is much less profound than the
exchange hole especially at high density. Whereas the exchange hole goes to zero at the position
of the given electron and totally must contain one electron (meaning that Coulomb repulsion
must exclude one electronic charge), the correlation hole integrates up to zero,
i.e. it represents just a redistribution of charge around a given electron. Furthermore there
is no condition for its amplitude at the site of the first electron.
The LDA formalism is applied including the corrections due to correlation and spin-polarization,
 and it has proven to
be very successful in electronic structure calculations \cite{ldas}. More recent corrections due to
weak variations of the density, such as in the generalized gradient approximation (GGA) have
improved some results of LDA without making the method more difficult \cite{gga}.

In this work we propose a different derivation of the DF-type potential for exchange and correlation
than what is done in HFS and DF theories. The correlation part between electrons of opposite spin
has been presented previously in ref. \cite{bar}. Here we adapt the formalism to densities of
equal spin, i.e. for the dominating part of the potential which is denoted exchange.
The term "exchange" is no longer appropriate, since as we shall see there is only correlation
due to Coulomb repulsion between electrons. But there is an important boundary condition
which modifies completely the correlation hole for equal or opposite spin. However, the term "exchange"
is so common that we in the following continue to use it for the interaction
between electrons of equal spin. The goal is to  generalize the method
for non-constant densities as in real atoms and solids, but in this work we
consider free-electron conditions of constant density.

The electron gas parameter $r_s$ determines the average radius around a given electron
of density $\rho$, $4\pi/3 r_s^3 \rho = 1$. An electron at some point will on the average
have another electron $\sim r_s$ away from it, so the Coulomb repulsion is of the order
$1/r_s$ when using atomic units. The density in a solid is large near to the atoms and
$r_s$ is small and rarely larger than 2 a.u., i.e. always smaller than the atomic radius. In the near
neighborhood of an electron, which is fixed at some position $r=0$, 
other electrons will feel a large repulsive Coulomb potential due to the fixed electron.
The amplitude of this potential will even be larger than the mean-field crystal potential
that determines the crystal wave function $\Psi_i(r)$ in eq
\ref{eq:sch} \cite{ust}.  $\Psi_i(r)$ will not be a good description for the wave function of other
electrons near the one fixed at $r=0$. This is even more so for electrons of equal spin,
where a second electron cannot be at the origin. Therefore, in order to determine the
exchange and correlation potential one must consider that $\Psi_i(r)$ is deformed by the Coulomb
interaction and the boundary condition for equal spin,
around the single electron that we temporarily fix at a local origin. This happens for all 
surrounding electrons, independent of which shell they belong to, when the Coulomb repulsion
excedes the value of the crystal potential at the local origin. For equal spin, this 
is even more certain, since the boundary condition for r=0 apply to all electrons independent
of their relative differences in kinetic energy.

First, for electrons of opposite spin and total density $\rho/2$ (no spin-polarization)
the redistributed charge density can be determined from a Schr\"{o}dinger equation
of type
\begin{equation}
\label{eq:corr}
-\nabla^2 \phi(r) + (g/r + \int \frac{\rho(r')}{\mid r - r' \mid}d^3r' + 
\mu_{xc}(r))\phi(r) = \epsilon \phi(r)
\end{equation}
with the boundary condition that the density $\phi(r)^2$ should tend to $\rho/2$ at the
limit of the correlation hole (at $r=r_s$ or so) \cite{bar}. In eq. \ref{eq:corr}, $g$ is the effective
interaction strength and the following two potential terms are due to the possible Coulomb
exchange-correlation interaction among more than one electron within the correlation hole. (It is
not necessary to limit the problem to two electrons, one at $r=0$ and the other around it).
Further in eq. \ref{eq:corr} the effective mass is 1/2. The lowest energy $\epsilon$ is for
$\phi(r)$ being an $s$-state, i.e. $\ell=0$.

Secondly, for electrons of the same spin as the one at the center, one can solve a similar
type of equation. However, we need to recall the Pauli principle a second time:
Two electrons of the same spin cannot be in the same state, and in particular at a given
instant they cannot be at the same place. This puts a boundary condition on the wave function
$\varphi(r)$ for electrons of the same spin; $\varphi(0)$=0, and it implies solutions of $\ell \geq 1$.
Furthermore, the electron at the center is unique and cannot be contained in the surrounding
cloud,

\begin{equation}
\label{eq:norm}
\int (\varphi(r)^2 - \rho/2)d^3r = 1   
\end{equation}

Since correlation is allowed even between electrons of the same spin, it is possible that
$\varphi(r)^2$ could be larger than $\rho/2$ for some points far from 0, but at the limit
of the exchange hole $\varphi(r)^2$ should tend $\rho/2$ as in case of correlation only.
In principle, this requires a continous matching of both the amplitude and the derivative, 
but in this work only one continuity condition is applied. The equation for $\varphi$ becomes;
\begin{equation}
\label{eq:ex}
-\nabla^2 \varphi(r) + (g/r + \ell(\ell+1)/r^2 +\int \frac{\rho(r')}{\mid r - r' \mid}d^3r' + 
\mu_{xc}(r))\varphi(r) = \epsilon \varphi(r)
\end{equation}

with $\ell$=1 for the lowest energy. The influence of the coupling strength $g$ is strong
on $\phi(r)$, but only minor on $\varphi(r)$.

The DF value for the xc-energy is 
\begin{equation}
\varepsilon = 0.5 \int \int (\varphi_g(r)-\rho/2)/r d^3r dg
\end{equation}
and 
\begin{equation}
\varepsilon = 0.5 \int \int (\phi_g(r)-\rho/2)/r d^3r dg 
\end{equation}
for equal and opposite spin, respectively.
The integration over the coupling strength is from 0 to 1, and the potential is found from \cite{lda,ldas}
\begin{equation}
\mu = \frac{d}{d\rho}(\rho \varepsilon)  
\end{equation}

These equations are solved for different densities, but some technical points are to be noted.
The boundary condition for $\varphi(r_c)$ is that the derivative at $r_c$ is zero, where
$4\pi/3 r_c^3 \rho = 4$. This makes approximately $\varphi^2(r_c) = \rho/2$. Smaller $r_c$ gives a very
similar $\mu$-potential, but the discontinuity at the hole boundary is evident. The same $r_c$ is
used for the correlation between opposite spins with the boundary condition  $\phi^2(r_c) = \rho/2$.
$\phi(r)$ has one maximum (larger than $\rho$) before $r=r_c$, while $\varphi(r)$ is an ever
increasing function. The results for the coupling strength $g$=1/2 are very close to those
with the integrated values from 0 to 1. Finally, we ignore the coupling between the two
spin densities since it turns out to have a quite small effect on the result
(even ignoring terms 3 and 4 in eqs. \ref{eq:corr} and \ref{eq:ex}
give almost the same result). 

The high density solution for equal spin $\varphi(r)$ with $g$=0, gives a potential value which is
about 12 \% larger than the KS-value. 
The reason for this is probably connected with
the imposed cut-off radius. The Slater function \cite{sla}, which defines the hole in KS-theory, has
Friedel-like oscillations outside its first node.  A simple exercise will motivate that the present
scheme should give the KS-value for interaction $g$=0, if the solutions for large $r$ are retained.
For very large $r$ it is simpler to handle solutions going to zero than solutions which are approaching
the electron density. Therefore we search the solutions $j(r)$ for a fictive positive charge that cancels
exactly one electron, and which equals the electron density at $r$=0. Instead of eq. \ref{eq:ex}, we have
\begin{equation}
-\nabla^2 j(r) + \ell(\ell+1)/r^2 j(r) = \epsilon j(r)
\end{equation}
when no interaction within the electron cloud is considered. This is the V=0 central potential
problem with the known solutions, the Bessel functions $j_{\ell}(\sqrt{\epsilon}r)$.
The solution $j_0$ fulfills the condition at $r$=0, but it is not normalizable over all space.
Linear combinations with higher-$\ell$ functions fulfill the condition at $r$=0, but $(j_0 + j_1)$
is not normalizable either. The low-$\ell$ combination which fulfills the two conditions is
$(j_0 + j_2)$. The square of this solution is precisely the Slater function, normalized to one electron: 
\begin{equation}
\int^{\infty}_0 (j_0+j_2)^2 d^3r = 1
\end{equation}

At the first node of $(j_0 + j_2)$ the normalization is about 0.8, and the potential is
about 95 \% of its full value, the KS-value. Verification by  numerical solutions will be difficult
because of the requirement of normalization over all space. But at the same time 
it would be incorrect to impose a cut-off at the first node, as it will
localize the hole too much. Instead we can return to the solutions for the electron density ($\varphi^2(r)$
instead of the fictive positive density) and extend the solution beyond the first maximum. By letting the interacting
density reach the non-interacting density at $r_c$, and normalizing to unity at a larger radius, one finds that
some fraction of the hole is beyond the maximum at $r_c$, similar to (but still more localized than) 
the Slater function. This procedure
is used with the coupling $g$, although it is no longer required that $\varphi^2 \equiv \rho/2$ at some point,
because of the Coulomb repulsion.  
One can summarize the final results in terms of powers of the 
density,

\begin{equation}
\mu = 2.1 \rho^{1/3} + 0.07 \rho^{1/6}  (Ryd)
\end{equation}
where the major part of the second term is the part from different spin. The first part is still
a little larger (6 \%)
than the KS-value for exchange, $1.961 \rho^{1/3}$ $(Ryd)$. The effective parameter ($\alpha$)
in front of the $\rho^{1/3}$-term is shown in fig 1, together with two commonly used LDA potentials. 

In the case of spin polarization, it is possible to solve the equations above for two different
densities. Due to the Hartree and $\mu_{xc}$ terms in eqns. \ref{eq:corr} and \ref{eq:ex}, there
is a coupling between the two densities, which diminish the tendency for polarization. However,
this interaction will also complicate the search for self-consistent solutions, and approximate
results can be obtained rather quickly by omitting the two terms completely.  The preliminary results 
give a slightly larger tendency for polarization than standard spin-polarized versions of LDA \cite{ldas,bh}. 

 The resulting potential is close to, but not better than 
LDA. However, the results are
suffiently promising to continue the search for improved solutions at large $r$ and for non-locality. 
Apart from the problem at large $r$ the formalism contains no free parameters or other ad-hoc assumptions.
Further, the present formalism helps to understand one difference between HF and DF approaches.
The exchange in HF acts selectively between orbitals, so the exchange between core and valence
appears different from within valence states, for example. In LDA all electrons have the same status
independent of the electronic shell, since the density is made up by all electrons, core and valence.
(However, in practise HF and LDA results may come out not too different after convergence.) From the way
we look at the electron interactions in the present method, it is natural that all electrons should
be equal because of the boundary condition at $r$=0. Equal spin electrons have zero probability to
be at the same place, and this is independent of the energy of their atomic orbital. For opposite
spin electrons the probablity is not zero but significanly reduced, because the Coulomb repulsion
is strong and will separate the electrons, independently of their kinetic energy. Selective
shell dependent interactions like a self-interaction correction, is not a natural extension
of the present approach.
 Also methods which assign a 
strong orbital dependent parameter like a Hubbard parameter to some of the electrons, are difficult
to understand, because the electron cannot distinguish between the interaction to a special orbital
from the rest of the electron density. However, all this reasoning has to be revised if the 
interaction length ($r_c$ roughly), becomes large in comparison to the size of an atom. A large
part of the core density may exist as a "bump" within the radius $r_c$, and corrections
due to non-locality will be important. Therefore it is essential to correct the LDA as in GGA \cite{gga}
or for even stronger non-constant density corrections, especially for f-electron systems where
self-interaction or Hubbard models give different results than LDA.

 In conclusion, the most important result is that
the hole for equal spins has similar shape as the exchange hole for free electrons. As a cut-off
radius is imposed, it will be slightly more localized and give a slightly stronger potential than
the exchange hole which has small oscillations far away. The zero amplitude at the origin is
the result of a boundary condition, and it is the rigidity of the Schr\"{o}dinger
equation which determines the shape of the exchange hole for equal spins. Further development of
the long-range tails of the hole is necessary before applications, and generalization for
non-constant density will be of great interest.




\begin{figure}
\epsfxsize=300pt
\epsffile{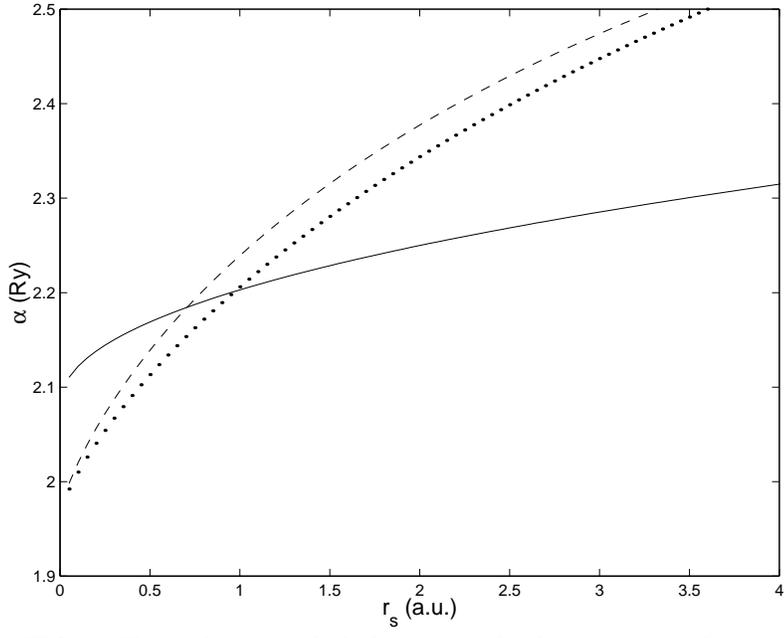}
\caption{The prefactor $\alpha$ which determines the density potential according to
$\mu_{xc} = \alpha \rho^{1/3}$. The broken and dotted lines are from refs. \protect\cite{ldas} and \protect\cite{sstl}, respectively.}
\label{fig1}
\end{figure}

\end{document}